\documentstyle{article}
\begin{document}
\title {
\bf \huge Bethe Ansatz for the $SU(4)$ Extension of the Hubbard Model
}

\author{
{\bf
Heng Fan\thanks {
Permanent address: Institute of Modern Physics, 
Northwest University, Xi'an 710069, P.R.China
},  Miki Wadati}\\
\normalsize
Department of Physics, Graduate School of Science,\\
\normalsize University of Tokyo, Hongo 7-3-1,
Bunkyo-ku, Tokyo 113-0033, Japan.\\
}
\maketitle

\begin{abstract}
We apply the nested algebraic Bethe ansatz method to solve the eigenvalue
problem for the SU(4) extension of the Hubbard model. The Hamiltonian
is equivalent to the SU(4) graded permutation operator. The graded
Yang-Baxter equation and the graded Quantum Inverse Scattering Method
are used to obtain the eigenvalue of the SU(4) extension of the
Hubbard model.
\end{abstract}              
       


\baselineskip 0.5truecm
\section{Introduction}
Since the discovery of high-temperature superconductivity, much more
attention has been paied to the theoretical mechanism for such
phenomena. Most proposals concern about the Hubbard model and the
$t-J$ model \cite{AZR}. C.N.Yang \cite{Yang} advocated the importance
of $\eta$-pairing mechanism and the property of off-diagonal long-range
order (ODLRO) for the eigenfunctions in superconductivity.
And it is stated that the wave function of BCS theory do have the ODLRO.
Essler, Korepin and Schoutens \cite{EKS} proposed an $SU(2|2)$
extended Hubbard model with the
$\eta$-pairing symmetry. They showed that the eigenstates of
this extended Hubbard model exhibit ODLRO and is thus superconducting.
A more general extension of the Hubbard model with $\eta $-pairing
is also proposed in ref.\cite {BKS}, see
ref.\cite{FLR} for some related works.
One of the authors showed that the $SU(2|2)$ Hamiltonian can be
constructed by the coupled $SU(1|1)$ chains, and
proposed a new $SU(4)$ extension of the Hubbard model\cite{F}.
It can be proved that the eigenstate which shows ODLRO of
the $SU(2|2)$ case is also an eigenstate for 
the $SU(4)$ extension of the Hubbard model. Thus we present another
extension of the Hubbard model, its eigenstate possesses the ODLRO property.
The $SU(4)$ extension of the Hubbard model is constructed by two
coupled $SU(2)$ chains. To some extent, it is connected with the
spin-ladders which are introduced to describe the quasi-one-dimensional
materials in the superconductivity\cite{DR}.
Recently, a series of exact solvable spin-ladders including coupled
$t-J$ model were proposed\cite{NT}. Different from the
$SU(4)$ spin-ladder model, the $SU(4)$ extension of the Hubbard
model has the Hubbard-like interactions instead of the rung interactions
in the ladder models.
And another
motivation for us to study the $SU(4)$ extension of the Hubbard model
is that it is the fermion version of the $SU(4)$ spin chain with
orbital degeneracy\cite{LMSZ,LMSF1}.

In this paper, we shall give a detailed calculation for the eigenvalues
of the $SU(4)$ extension of the Hubbard model. This extension of
the Hubbard model is a fermion chain, so we naturally introduce the graded
Quantum Inverse Scattering Method (QISM) to solve the eigenvalue problem.
The Hamiltonian is presented to be an $SU(4)$ graded permutation
operator. We thus start from the rational $SU(4)$ R-matrixt.
By using the graded method, we show that the Hamiltonian can
be obtained from the transfer matrix with periodic boundary
conditions. Then, we can apply the graded nested Bethe ansatz method to
find exactly the eigenvalues of the Hamiltonain.

The paper is organized as follows:
We shall introduce our model and fix notations in the next section. We then
shall prove that the one-dimensional Hamiltonian is equal to
the graded permutation operator and the system is integrable in
section 3. In section 4,
the graded algebraic Bethe ansatz method will be used to obtain the
eigenvalue of the Hamiltonian. the corresponding Bethe ansatz equations
will be presented. Finally,
we give a brief analysis of the Bethe ansatz equations in the last section.

\section{Description of the Model}
The Hamiltonian of the $SU(4)$ extension of the Hubbard model is
written as:
\begin{eqnarray}
H=-H^0+U\sum _{j=1}^N(n_{j,\uparrow }-{1\over 2})
(n_{j,\downarrow }-{1\over 2}),
\label{Ham}
\end{eqnarray}
where $H^0=\sum _{jk}H^0_{jk}$
is the two coupled $SU(2)$ fermion chains, and the summation takes for
nearest-neighbour sites. The local Hamiltonian of the
coupled $SU(2)$ fermion chain has
the form
\begin{eqnarray}
H^0_{jk}&=&[c_{j\uparrow}^{\dagger }c_{k\uparrow}
+c_{k\uparrow}^{\dagger }c_{j\uparrow}
+2n_{j\uparrow}n_{k\uparrow}
-n_{j\uparrow}
-n_{k\uparrow}+1]
\nonumber \\
&&\times   
[c_{j\downarrow}^{\dagger }c_{k\downarrow}+
c_{k\downarrow}^{\dagger }c_{j\downarrow}
+2n_{j\downarrow}n_{k\downarrow}
-n_{j\downarrow}
-n_{k\downarrow}+1].
\label{local}
\end{eqnarray}
The operator $c_{j\sigma}$ and $c_{j\sigma }^{\dagger }$ represent the
annihilation and creation operators of electrons with spin
$\sigma =\uparrow ,\downarrow $ on
a lattice site $j$. These operators are canonical Fermion
operators satisfying anticommutation relations
expressed as
\begin{eqnarray}
\{c_{i\sigma}^{\dagger}, c_{j\tau }\} =\delta _{ij}
\delta _{\sigma \tau },
~~\{c_{i\sigma}, c_{j\tau }\}
=\{c_{i\sigma}^{\dagger}, c_{j\tau }^{\dagger}\} =0.
\end{eqnarray}
We denote by $n_{j\sigma }=c_{j\sigma }^{\dagger }c_{j\sigma }$ the
number operator for the electrons on a site $j$ with spin $\sigma $.

Explicitly, the local Hamiltonian (\ref{local}) can be written as
\begin{eqnarray}
H^0_{jk}&=&\sum _{\sigma =\uparrow ,\downarrow }
[(c_{j\sigma}^{\dagger }c_{k\sigma}+
c_{k\sigma}^{\dagger }c_{j\sigma})
(1-n_{j,-\sigma})(1-n_{k,-\sigma})
+(c_{j\sigma}^{\dagger }c_{k\sigma}+
c_{k\sigma}^{\dagger }c_{j\sigma})n_{j,-\sigma}n_{k,-\sigma}]
\nonumber \\
&&+
c_{j\uparrow}^{\dagger }c_{j\downarrow}^{\dagger}      
c_{k\downarrow}c_{k\uparrow }
+
c_{j\downarrow}c_{j\uparrow }
c_{k\uparrow}^{\dagger }c_{k\downarrow}^{\dagger}
-c_{j\downarrow}^{\dagger }c_{j\uparrow}
c_{k\uparrow}^{\dagger }c_{k\downarrow }
-c_{j\uparrow}^{\dagger }c_{j\downarrow}
c_{k\downarrow}^{\dagger }c_{k\uparrow }
\nonumber \\
&&+{1\over 2}(n_{j\uparrow}-n_{j\downarrow})
(n_{k\uparrow}-n_{k\downarrow})
+{1\over 2}(n_j-1)(n_k-1)
\nonumber \\
&&+4(n_{j\uparrow }-{1\over 2})
(n_{j\downarrow }-{1\over 2})(n_{k\uparrow }-{1\over 2})
(n_{k\downarrow }-{1\over 2})+{1\over 4},
\end{eqnarray}
where we denote by $n_j=\sum _{\sigma =\uparrow ,\downarrow }n_{j\sigma }$
the number operator for the electrons on a site $j$.
The Hamiltonian $H^0$ is invariant under spin-reflection
$c_{j\uparrow}\leftrightarrow c_{j\downarrow }$. But it does
not have the invariant property under particle-hole replacement
$c^{\dagger }_{j\sigma }\leftrightarrow c_{j\sigma}$.
There are four kinds of state at a given site:
\begin{eqnarray}
|0>_j, |\uparrow >_j=c^{\dagger }_{j\uparrow}|0>_j,
|\downarrow >_j=c^{\dagger }_{j\downarrow}|0>_j,
|\uparrow \downarrow >_j=
c^{\dagger }_{j\downarrow}c^{\dagger }_{j\uparrow}|0>,
\end{eqnarray}
two of them are fermionic and the other two are bosonic.  The state
$|\uparrow \downarrow >$
represents that an electron-pair is localized on a single lattice site.
The formation of such pairs,
called localons, is considered to be
a mechanism to create `Cooper pairs'\cite{EKS}.

We introduce here some generators. The spin operators $S=\sum _{j=1}^NS_j$,
$S^{\dagger }$ and $S^z$,
\begin{eqnarray}
S_j=c_{j\uparrow }^{\dagger }c_{j\downarrow },
~~S_j^{\dagger }=c_{j\downarrow }^{\dagger }c_{j\uparrow },
~~S_j^z={1\over 2}(n_{j\uparrow }-n_{j\downarrow }),
\end{eqnarray}
form an $SU(2)$ algebra
$[S,S^{\dagger }]=2S^z, [S^{\dagger},S^z]=S^{\dagger},
[S,S^z]=-S$. Since the fermion operator is grassmann odd,
we find that the above spin operators are grassmann even (bosonic).
The $\eta $-pairing generators also form an $SU(2)$ algebra
$[\eta ,{\eta }^{\dagger }]=2\eta ^z,
[\eta ^{\dagger},\eta ^z]=\eta ^{\dagger},
[\eta ,\eta ^z]=-\eta $ with
\begin{eqnarray}
\eta _j=c_{j\uparrow }c_{j\downarrow },
~~\eta _j^{\dagger }=c_{j\downarrow }^{\dagger }c_{j\uparrow }^{\dagger },
~~\eta _j^z=-{1\over 2}n_j+{1\over 2}.
\end{eqnarray}
We see that those operators be grassmann even. The generator
$X_j=(n_{j\uparrow }-{1\over 2})(n_{j\downarrow }-{1\over 2})$
is also a grassmann even operator. We further introduce eight grassmann
odd generators. They take the following form:
\begin{eqnarray}
&&Q_{j\sigma }=(1-n_{j,-\sigma })c_{j\sigma},
~~Q_{j\sigma }^{\dagger }=(1-n_{j,-\sigma })c_{j\sigma}^{\dagger},
\nonumber \\
&&\tilde {Q}_{j\sigma }=n_{j,-\sigma }c_{j\sigma},
~~\tilde {Q}_{j\sigma }^{\dagger}=n_{j,-\sigma }c_{j\sigma}^{\dagger},
\end{eqnarray}
with $\sigma =\uparrow ,\downarrow $ representing
spin up and spin down respectively.
We shall focus in this paper on the one-dimensional case with
periodic boundary conditions, that means
the nearest-neighbour
sites of site $j$ are $j\pm 1$.
In terms of these generators, the one-dimensional Hamiltonian
$H^0_{jj+1}$ is written as
\begin{eqnarray}
H^0_{jj+1}&=&\sum _{\sigma =\uparrow,\downarrow}
[Q_{j\sigma}^{\dagger}Q_{j+1\sigma}+
Q_{j+1\sigma}^{\dagger}Q_{j\sigma}
+\tilde {Q}_{j\sigma}^{\dagger}\tilde {Q}_{j+1\sigma}+
\tilde {Q}_{j+1\sigma}^{\dagger}\tilde {Q}_{j\sigma}]
\nonumber \\
&&+\eta _j^{\dagger}\eta _{j+1}
+\eta _j\eta _{j+1}^{\dagger }+2\eta _j^z\eta _{j+1}^z
-S_j^{\dagger }S_{j+1}-S_jS_{j+1}^{\dagger }
+2S_j^zS_{j+1}^z+4X_jX_{j+1}+{1\over 4}.
\end{eqnarray}
Next, we shall show that the Hamiltonian $H^0_{jj+1}$
is equal to the graded $SU(4)$
permutation operator. Because it is the graded version
of the permutation operator, note that
this Hamiltonian does not preserve the
property of $SU(4)$ invariant, for example, we do not have relation
$[H^0, S]=0$.

We can use $4\times 4$ matrices to express the above generators.
The represention can be realized as
\begin{eqnarray}
&&S_j=E_j^{21}, ~~S_j^{\dagger }=E_j^{12},
~~S_j^z={1\over 2}(E_j^{22}-E_j^{11}),
\nonumber \\
&&\eta _j=E_j^{34}, ~~\eta _j^{\dagger }=E_j^{43},
~~\eta _j^z={1\over 2}(E_j^{33}-E_j^{44}),
\nonumber \\
&&Q_{j\uparrow }=E_j^{32}, ~~Q_{j\uparrow }^{\dagger }=E_j^{23},
~~Q_{j\downarrow }=E_j^{31},~~Q_{j\downarrow }^{\dagger }=E_j^{13},
\nonumber \\
&&\tilde {Q}_{j\uparrow }=-E_j^{14},
~~\tilde {Q}_{j\uparrow }^{\dagger }=-E_j^{41},
~~\tilde {Q}_{j\downarrow }=E_j^{24},
~~\tilde {Q}_{j\downarrow }^{\dagger }=E_j^{42},
\label{rep}
\end{eqnarray}
where the matix $E_j^{\alpha \beta }$ acts on the $j$-th space with
its elements defined as $(E_j^{\alpha \beta })_{\gamma \nu }=
\delta _{\alpha \gamma }\delta _{\beta \nu}$. We remark
that the matrix $E_j^{\alpha \beta }$ is not a conventional matrix
but a supermatrix with
grassmann number $\epsilon _{\alpha }+\epsilon _{\beta }$,
where $\epsilon _{\alpha }=0$ represents grassmann even (boson),
and $\epsilon _{\alpha }=1$ represents grassmann odd (fermion).
From the representations (\ref{rep}), we find that we should
choose the grading
$\epsilon _1=\epsilon _2=1,~~\epsilon _3=\epsilon _4=0$ so that the
representations have correct grassmann number.

Considering the above representation,
we find that the Hamiltonian can be written as 
\begin{eqnarray}
H_{jj+1}^0=\sum _{\alpha \not= \beta }
(-1)^{\epsilon _{\beta }}E_j^{\alpha \beta }
\otimes E_{j+1}^{\beta \alpha }+\sum _{\alpha }E_j^{\alpha \alpha }
\otimes E_{j+1}^{\alpha \alpha }.
\end{eqnarray}
The right hand side the this relation is just the graded $SU(4)$
permutation operator.
Now, we define the super (graded) tensor-product as
\begin{eqnarray}
[A\otimes B]_{mj,kl}=(-1)^{(\epsilon _m+\epsilon _k)\epsilon _j}A_{mk}B_{jl}.
\label{sup}
\end{eqnarray}
According to this definition,
the non-zero elements of the local Hamiltonian read as:
\begin{eqnarray}
[H_{jj+1}^0]_{\alpha \beta }^{\gamma \nu}=
\left\{ \begin{array}{l}
(-1)^{\epsilon _{\alpha }\epsilon _{\beta }}\delta _{\alpha \nu }
\delta _{\beta \gamma },~~\alpha \not= \beta ,\\
1, ~~~\alpha =\beta =\gamma =\nu .
\end{array}\right.
\label{ham}
\end{eqnarray}

Thus we have proved that the Hamiltonian $H^0$ is equal to the
graded permutation operator, and because the Hubbard interaction
term $U\sum _jX_j$ commute with $H^0$, they have common eigenvectors.
We then can use graded algebraic Bethe ansatz method to find the
eigenvalue of the Hamiltonian (\ref{Ham}).
We point out that the Hamiltonian $H^0$ is neither the usual $SU(4)$
chain nor the super $SU(2|2)$ chain. 

\section{The Integrability of the Model}
We begin with the $SU(4)$ rational R-matrix. The non-zero elements
of the matrix have the form
\begin{eqnarray}
\tilde {R}(u)_{aa}^{aa}&=&u+i,\nonumber \\
\tilde {R}(u)_{ab}^{ab}&=&(-1)^{\epsilon _a\epsilon _b}u,
~~a\not= b,\nonumber \\
\tilde {R}(u)_{ab}^{ba}&=&i, ~~a\not =b,
\end{eqnarray}
where indices $a,b$ take values $1,\cdots,4$, $i=\sqrt {-1}$ is the
cross parameter.  We know this R-matrix satisfies
the usual Yang-Baxter equation                           
\begin{eqnarray}
\tilde {R}_{12}(u_1-u_2)\tilde {R}_{13}(u_1-u_3)\tilde {R}_{23}(u_2-u_3)
= \tilde {R}_{23}(u_2-u_3)\tilde {R}_{13}(u_1-u_3)\tilde {R}_{12}(u_1-u_2).
\end{eqnarray}
As in section 2, we still choose the grading
$\epsilon _1=\epsilon _2=1, ~\epsilon _3=\epsilon _4=0$, that means
the grading is FFBB.
Introducing a diagonal matrix
$I_{ac}^{bd}=(-1)^{\epsilon _a\epsilon _c}\delta _{ab}\delta _{cd}$,
we modify the original R-matrix to the following form
\begin{eqnarray}
R(u)=I\tilde {R}(u).
\end{eqnarray}
For the non-zero elements of the R-matrix $R_{ab}^{cd}$, we have
a relation
$\epsilon _a+\epsilon _b+\epsilon _c+\epsilon _d=0$. Thus we find the
new defined R-matrix satisfies the graded Yang-Baxter equation which
reads explicitely as
\begin{eqnarray}
R(u-v)_{a_1a_2}^{b_1b_2}
R(u)_{b_1a_3}^{c_1b_3}
R(v)_{b_2b_3}^{c_2c_3}
(-)^{(\epsilon _{b_1}+\epsilon _{c_1})\epsilon _{b_2}}
=
R(v)_{a_2a_3}^{b_2b_3}R(u)_{a_1b_3}^{b_1c_3}
R(u-v)_{b_1b_2}^{c_1c_2}(-)^{(\epsilon _{a_1}
+\epsilon _{b_1})\epsilon _{b_2}}.
\label{gybe}
\end{eqnarray}
We remark that, we have already graded the R-matrix, that means
the matrix is now a supermatrix with grassmann numbers.
For $u=0$, $R_{12}(0)=iP_{12}$,
and the elements of $P_{12}$ read
$[P_{12}]_{ab}^{cd}=(-1)^{\epsilon _a\epsilon _b}\delta _{ad}\delta _{bc}$.
We know that this $P_{12}$
is the super permuation operator corresponding to group $SU(2|2)$ with
FFBB grading.

In the framework of the QISM, we can construct
the $L$ operator from the R-matrix as:
\begin{eqnarray}
L_{aq}(u)\equiv R_{aq}(u),
\end{eqnarray}
where $a$ represents the auxiliary space and $q$
represents the quantum space.  
Thus we have the graded Yang-Baxter relation
\begin{eqnarray}
R_{12}(u-v)L_1(u)L_2(v)
=L_2(v)L_1(u)R_{12}(u-v).
\label{ybr}
\end{eqnarray}
Here the tensor product is understood to be
in the sense of super tensor product
defined in relation (\ref{sup}).
In the rest of this paper, all tensor products are in the super sense.

Following the standard method of QISM, the row-to-row monodromy matrix
$T_N(u)$ is defined as the matrix product over
the $N$ operators on all sites of the lattice, 
\begin{eqnarray}
T_a(u)=L_{aN}(u)L_{aN-1}(u)\cdots L_{a1}(u),
\label{mon}
\end{eqnarray}
where $a$ still represents the auxiliary space.
Explicitely we write
\begin{eqnarray}
&&\{ [T(u)]^{ab}\}_{\begin {array}{c}
\alpha _1\cdots \alpha _N\\
\beta _1\cdots \beta _N\end{array}}
\nonumber \\
&=&L_N(u)_{a\alpha _N}^{c_N\beta _N}
L_{N-1}(u)_{c_N\alpha _{N-1}}^{c_{N-1}\beta _{N-1}}
\cdots L_1(u)_{c_2\alpha _1}^{b\beta _1}
(-1)^{\sum _{j=2}^N(\epsilon _{\alpha _j}+\epsilon _{\beta _j})
\sum _{i=1}^{j-1}\epsilon _{\alpha _i}}
\label{grad}
\end{eqnarray}
By repeatedly using the Yang-Baxter relation (\ref{ybr}),
one can prove that
the monodromy matrix also satisfies the Yang-Baxter relation
\begin{eqnarray}
R_{12}(u-v)T_1(u)T_2(v)
=T_2(v)T_1(u)R_{12}(u-v).
\label{YBR}
\end{eqnarray}

We consider in this paper the periodic boundary condition.
The transfer matrix $t(u)$
of this model is defined as the supertrace of the
monodromy matrix in the auxiliary space. In general case,
the supertrace is defined as
\begin{eqnarray}
t (u)=strT(u)                     
=\sum (-1)^{\epsilon _a}T(u)_{aa}.
\end{eqnarray}

As a consequence of the Yang-Baxter relation (\ref{YBR}) and the unitarity
property of the R-matrix
\begin{eqnarray}
R_{12}(u)R_{21}(-u)=(i+u)(i-u)\cdot id.,
\end{eqnarray}
we can prove that
the transfer matrix commutes each other for different spectral
parameters.
\begin{eqnarray}
[t (u),t (v)]=0 
\end{eqnarray}
Generally in this sense the model is
integrable. Expanding the transfer matrix in the powers of $u$,
we can find conserved quantites, the first non-trivial conserved
equantity is the Hamiltonian.

For the R-matrix under consideration,
the Hamiltonian can be obtained by taking the 
logarithmic derivative of the transfer matrix at the zero spectral parameter, 
\begin{eqnarray}
H'\equiv \sum _{j}^NH'_{jj+1}
=\frac {d\ln [t (u)]}{du}|_{u=0}
=\sum _{j}^NP_{jj+1}L'_{jj+1}(0).
\end{eqnarray}
Explicitly, we write the local Hamiltonian $H'_{jj+1}$ obtained
above in a matrix form
\begin{eqnarray}
H'_{jj+1}=\left( \begin{array}{cccc}
E_{j+1}^{11}& -E_{j+1}^{21}& E_{j+1}^{31}& E_{j+1}^{41}\\  
-E_{j+1}^{12}& E_{j+1}^{22}& E_{j+1}^{32}& E_{j+1}^{42}\\
E_{j+1}^{13}& E_{j+1}^{23}& E_{j+1}^{33}& E_{j+1}^{43}\\
E_{j+1}^{14}& E_{j+1}^{24}& E_{j+1}^{34}& E_{j+1}^{44}\end{array}\right),
\end{eqnarray}
where $E^{ab}_{j+1}$ acts on $j+1$-th space. We can find this Hamiltonian
is exactly the Hamiltonian $H^0_{jj+1}$ (\ref{ham}).
We show explicitely here what we mentioned above,
this Hamiltonian is not
the usual $SU(4)$ Hamiltonian because we have negative signs
in the above relation,
it is also different from the Hamiltonian of the supergroup $SU(2|2)$
for all diagonal elements are positive here.
We have proved the Hamiltonian under consideration        
can be obtained from the transfer matrix of the exactly solvable model
and therefore is integrable.

\section{Graded Algebraic Bethe Ansatz Method}
We use the nested algebraic Bethe ansatz method to find the
eigenvalues of the transfer matrix. We start from the graded rational
$SU(4)$ R-matrix with FFBB grading. Explicit form of the R-matrix is
\begin{eqnarray}
R(u)=\left( \begin{array}{cccccccccccccccc}
-u-i&0&0&0&0&0&0&0&0&0&0&0&0&0&0&0\\
0&u&0&0&-i&0&0&0&0&0&0&0&0&0&0&0\\
0&0&u&0&0&0&0&0&i&0&0&0&0&0&0&0\\
0&0&0&u&0&0&0&0&0&0&0&0&i&0&0&0\\
0&-i&0&0&u&0&0&0&0&0&0&0&0&0&0&0\\
0&0&0&0&0&-u-i&0&0&0&0&0&0&0&0&0&0\\
0&0&0&0&0&0&u&0&0&i&0&0&0&0&0&0\\
0&0&0&0&0&0&0&u&0&0&0&0&0&i&0&0\\
0&0&i&0&0&0&0&0&u&0&0&0&0&0&0&0\\
0&0&0&0&0&0&i&0&0&u&0&0&0&0&0&0\\
0&0&0&0&0&0&0&0&0&0&u+i&0&0&0&0&0\\
0&0&0&0&0&0&0&0&0&0&0&u&0&0&i&0\\
0&0&0&i&0&0&0&0&0&0&0&0&u&0&0&0\\
0&0&0&0&0&0&0&i&0&0&0&0&0&u&0&0\\
0&0&0&0&0&0&0&0&0&0&0&i&0&0&u&0\\
0&0&0&0&0&0&0&0&0&0&0&0&0&0&0&u+i
\end{array}\right).
\label{R}
\end{eqnarray}
This R-matrix satisfies the graded Yang-Baxter equation (\ref{gybe}). In
the framework of the QISM, the L-operator is defined by t
\begin{eqnarray}
L_j(u)=\left( \begin{array}{cccc}
u-b(u)E_j^{11}&-iE_j^{21}&iE_j^{31}&iE_j^{41}\\
-iE_j^{12}&u-b(u)E_j^{22}&iE_j^{32}&iE_j^{42}\\
iE_j^{13}&iE_j^{23}&u+iE_j^{33}&iE_j^{43}\\
iE_j^{14}&iE_j^{24}&iE_j^{34}&u+iE_j^{44}
\end{array}\right) ,
\end{eqnarray}
where $E_j^{\alpha \beta }$ acts on the $j$-th quantum space,
and we use notation $b(u)\equiv 2u+i$.
Considering the representations of the operator $E_j^{\alpha \beta }$
(\ref{rep}), we can also write L-operator in the following form

\begin{eqnarray}
&&L_j(u)=
\nonumber \\
&&\left( \begin{array}{cccc}
u-b(u)({1\over 4}-X_j-S_j^z)&-iS_j&iQ_{j\downarrow}
&-i\tilde {Q}_{j\uparrow }^{\dagger }\\
-iS_j^{\dagger }&u-b(u)({1\over 4}-X_j+S_j^z)&iQ_{j\uparrow}&
i\tilde {Q}_{j\downarrow }^{\dagger }\\
iQ_{j\downarrow }^{\dagger }
&iQ_{j\uparrow }^{\dagger }&u+i({1\over 4}+X_j+\eta _j^z)&
i\eta _j^{\dagger }\\
-i\tilde {Q}_{j\uparrow }&i\tilde {Q}_{j\downarrow }&
i\eta _j&u+i({1\over 4}+X_j-\eta _j^z)
\end{array}\right). 
\nonumber \\
\end{eqnarray}
We can also write it as
\begin{eqnarray}
&&L_j(u)=
\nonumber \\
&&
\left( \begin{array}{cccc}
u-b(u)(n_{j\downarrow }-n_{j\uparrow }n_{j\downarrow })
&-ic_{j\uparrow }^{\dagger }c_{j\downarrow }
&i(1-n_{j\uparrow })c_{j\downarrow}
&-i(1-n_{j\downarrow })c_{j\uparrow }^{\dagger }\\
-ic_{j\downarrow }^{\dagger }c_{j\uparrow }
&u-b(u)(n_{j\uparrow }-n_{j\uparrow }n_{j\downarrow })
&i(1-n_{j\downarrow })c_{j\uparrow }&
in_{j\uparrow }c_{j\downarrow }^{\dagger }\\
i(1-n_{j\uparrow })c_{j\downarrow }^{\dagger }
&i(1-n_{j\downarrow })c_{j\uparrow }^{\dagger }
&u+i(1-n_j+n_{j\uparrow }n_{j\downarrow })&
ic_{j\downarrow }^{\dagger }c_{j\uparrow }^{\dagger }\\
-in_{j\downarrow }c_{j\uparrow }&i
n_{j\uparrow }c_{j\downarrow }&
ic_{j\uparrow }c_{j\downarrow }&u+in_{j\uparrow }n_{j\downarrow }
\end{array}\right). 
\nonumber \\
\end{eqnarray}
We choose the local vacuum state as $|vac>_j=(0,0,0,1)^t$.
Actually this local vacuum state is equal to the state
$|\uparrow \downarrow >_j$.
Acting the L-operator on this local vacuum state, we have
\begin{eqnarray}
L_j(u)|vac>_j&=&\left(
\begin{array}{cccc}
u&0&0&0\\
0&u&0&0\\
0&0&u&0\\
-iE_j^{14}&iE_j^{24}&iE_j^{34}&u+i\end{array}\right) |vac>_j
\nonumber \\
&=&\left( \begin{array}{cccc}
u|\uparrow \downarrow >_j&0&0&0\\
0&u|\uparrow \downarrow >_j&0&0\\
0&0&u|\uparrow \downarrow >_j&0\\
i|\downarrow>_j&i|\uparrow >_j&i|0>_j&(u+i)|\uparrow \downarrow >_j
\end{array}
\right).
\label{lon}
\end{eqnarray}
Define the vacuum state as $|vac>=\otimes _{j=1}^N|vac>_j$. This vacuum
state is the full-filled state just like the `Dirac sea'. Using the
standard QISM, we denote the monodromy matrix defined in (\ref{mon}) as
follows,
\begin{eqnarray}
T(u)=\left( \begin{array}{cccc}
A_{11}(u)&A_{12}(u)&A_{13}(u)&B_1(u)\\
A_{21}(u)&A_{22}(u)&A_{23}(u)&B_2(u)\\
A_{31}(u)&A_{32}(u)&A_{33}(u)&B_3(u)\\
C_1(u)&C_2(u)&C_3(u)&D(u)\end{array}\right).
\end{eqnarray}
The transfer matrix with periodic boundary conditions is thus
written explicitely as
\begin{eqnarray}
t (u)=D(u)+A_{33}(u)-A_{22}(u)-A_{11}(u)
=D(u)+\sum _a(-1)^{\epsilon _a}A_{aa}(u).
\label{tran}
\end{eqnarray}
We write the last equation for convenience in using the nested Bethe
ansatz method later.
With the help of the definition of the monodromy matrix (\ref{mon})
and the result in relation (\ref{lon}), we find that
the action of the monodromy matrix on the vacuum state is
\begin{eqnarray}
T(u)|vac>
=\left( \begin{array}{cccc}
u^N &0 &0 &0\\
0&u^N&0&0\\
0&0&u^N&0\\
C_1(u)&C_2(u)&C_3(u)&(u+i)^N\end{array}\right) |vac>.
\label{act}
\end{eqnarray}
From the Yang-Baxter relation (\ref{YBR}), we can find the
commutation relations which are necessary for the algebraic Bethe ansatz
method,
\begin{eqnarray}
D(u)C_c(v)&=&\frac {v-u+i}{v-u}C_c(v)D(u)-\frac {i}{v-u}C_c(u)D(v),
\\
A_{ab}(u)C_c(v)&=&(-1)^{(\epsilon _a+\epsilon _b)\epsilon _c}
\frac {R^{(1)}(u-v)_{b_1d_1}^{bc}}{u-v}C_{d_1}(v)A_{ab_1}(u)
-(-1)^{\epsilon _a\epsilon _b}\frac {i}{u-v}C_b(u)A_{ac}(v),
\\
C_{c_1}(v)C_{c_2}(u)&=&(-)^{\epsilon _{b_1}\epsilon _{b_2}}
\frac {R^{(1)}(v-u)_{b_1b_2}^{c_1c_2}}{v-u+i}C_{b_2}(u)C_{b_1}(v).
\end{eqnarray}
Here all indices take values 1,2,3, and matix $R^{(1)}$ is equal to
the original R-matrix (\ref{R}) with indices being just 1,2,3.
We assume the eigenvectors of the transfer matrix as
\begin{eqnarray}
C_{c_1}(v_1)C_{c_2}(v_2)\cdots C_{c_m}(v_m)|vac>F^{c_1\cdots c_m},
\end{eqnarray}
where $F^{c_1\cdots c_m}$ is a function of the spectral parameters $v_j$.
Operating the transfer matrix (\ref{tran}) on this assumed eigenvector,
applying repeatedly the commutation relations and using the
result (\ref{act}), we have                              
\begin{eqnarray}
&&t (u)C_{c_1}(v_1)C_{c_2}(v_2)\cdots C_{c_m}(v_m)|vac>F^{c_1\cdots c_m},
\nonumber \\
&=&[D(u)+(-1)^{\epsilon _a}A_{aa}(u)]
C_{c_1}(v_1)C_{c_2}(v_2)\cdots C_{c_m}(v_m)|vac>F^{c_1\cdots c_m}
\nonumber \\
&=&
(u+i)^N\prod _{j=1}^m\left( \frac {v_j-u+i}{v_j-u}\right)
C_{c_1}(v_1)C_{c_2}(v_2)\cdots C_{c_m}(v_m)|vac>F^{c_1\cdots c_m}
\nonumber \\
&&+u^N\prod _{j=1}^m\left( \frac {1}{u-v_j}\right)
t ^{(1)}(u,\{v_k\} )^{c_1\cdots c_m}_{d_1\cdots d_m}
C_{d_1}(v_1)C_{d_2}(v_2)\cdots C_{d_m}(v_m)|vac>F^{c_1\cdots c_m}
\nonumber \\
&&+u.t.,
\end{eqnarray}                                            
where $u.t.$ means unwanted terms, and $t ^{(1)}$ is the nested
transfer matrix. The eigenvalue problem of $t (u)$
becomes now the eigenvalue problem of $t ^{(1)}$.
Denote the eigenvalues of $t $ and $t ^{(1)}$ by $\Lambda $
and $\Lambda ^{(1)}$. In order to cancel the unwanted terms,
we need the following Bethe ansatz equations
\begin{eqnarray}
\prod _{j=1}^m(v_k-v_j-i)(v_k+i)^N+v_k^N\Lambda ^{(1)}(v_k)=0,
~~~~k=1,2,\cdots ,m.
\end{eqnarray}
The nested transfer matrix is written explicitely as
\begin{eqnarray}
t ^{(1)}(u,\{v_k\} )^{c_1\cdots c_m}_{d_1\cdots d_m}
&=&(-1)^{\epsilon _a}R^{(1)}(u-v_1)_{b_1d_1}^{ac_1}
R^{(1)}(u-v_2)_{b_2d_2}^{b_1c_2}\cdots
R^{(1)}(u-v_m)_{ad_m}^{b_{m-1}c_m}
\nonumber \\
&&(-1)^{(\epsilon _a+\epsilon _{b_1})\epsilon _{d_1}
+(\epsilon _a+\epsilon _{b_2})\epsilon _{d_2}
+\cdots +(\epsilon _a+\epsilon _{b_{m-1}})\epsilon _{d_{m-1}}}.
\label{tcd}
\end{eqnarray}
For the non-zero elements of the $R^{(1)}(u)_{ab}^{cd}$, we still have
$\epsilon _a+\epsilon _b+\epsilon _c+\epsilon _d=0$, so
we can prove the following relation in (\ref{tcd}),
\begin{eqnarray}
{(\epsilon _a+\epsilon _{b_1})\epsilon _{d_1}
+(\epsilon _a+\epsilon _{b_2})\epsilon _{d_2}
+\cdots +(\epsilon _a+\epsilon _{b_{m-1}})\epsilon _{d_{m-1}}}
=\sum _{j=2}^m(\epsilon _{c_j}+\epsilon _{d_j})\sum _{k=1}^{j-1}
\epsilon _{d_k}
\end{eqnarray}
That means the $R^{(1)}$ matrices in the above relation is still in
the graded tensor product form which agrees with the definition
in (\ref{grad}).
And the above defined nested transfer matrix can be defined as
the supertrace on the auxiliary space for the reduced monodromy matrix
which satisfies the Yang-Baxter relation.
\begin{eqnarray}
&&t ^{(1)}(u,\{ v_k\} )=strT^{(1)}(u,\{ v_k\} )
\nonumber \\
&&=
(-1)^{\epsilon _a}R^{(1)}(u-v_1)_{b_1d_1}^{ac_1}
R^{(1)}(u-v_2)_{b_2d_2}^{b_1c_2}\cdots
R^{(1)}(u-v_m)_{ad_m}^{b_{m-1}c_m}
(-1)^{\sum _{j=2}^m(\epsilon _{c_j}+\epsilon _{d_j})\sum _{k=1}^{j-1}
\epsilon _{d_k}},
\end{eqnarray}
\begin{eqnarray}
R^{(1)}_{12}(u-v)T_1^{(1)}(u,\{ v_k\} )T_2^{(1)}(v,\{ v_k\} )
=T_2^{(1)}(v,\{ v_k\} )T_1^{(1)}(u,\{ v_k\} )R_{12}^{(1)}(u-v).
\label{YBR1}
\end{eqnarray}
Thus we deal with almost the same problem as the original one.
Repeating the graded algebraic Bethe ansatz method with grading FFB,
we can find
\begin{eqnarray}
\Lambda ^{(1)}(u)=\prod _{j=1}^{m^{(1)}}\left(
\frac {\mu _j-u+i}{\mu _j-u}\right)
\prod _{k=1}^m(u-v_k+i)+\prod _{j=1}^{m^{(1)}}
\left( \frac {1}{u-\mu _j}\right)
\prod _{k=1}^m(u-v_k)\Lambda ^{(2)}(u).
\end{eqnarray}
The eigenvalue $\Lambda ^{(2)}(u)$ for grading FF can be obtained
almost in the same way for the normal six-vertex model. We have
\begin{eqnarray}
\Lambda ^{(2)}(u)=-\prod _{j=1}^{m^{(2)}}
\left( \frac {\nu _j-u+i}{u-\nu _j}\right)
\prod _{k=1}^{m^{(1)}}(\mu _k-u-i)
-\prod _{j=1}^{m^{(2)}}\left(
\frac {\nu _j-u-i}{u-\nu _j}\right)
\prod _{k=1}^{m^{(1)}}(u-\mu _k).
\end{eqnarray}
In addition to these relations, we denote by $\Lambda (u)$ the
eigenvalue of the original transfer matrix $t(u)$. We have
\begin{eqnarray}
\Lambda (u)=\prod _{j=1}^m\left( \frac {v_j-u+i}
{v_j-u}\right) (u+i)^N+
\prod _{j=1}^m\left( \frac {1}{u-v_j}\right) u^N
\Lambda ^{(1)}(u),
\end{eqnarray}
where $v_j, \mu _j,\nu _j$ should satisfy their corresponding Bethe ansatz
equations. To summarize, the Bethe ansatz equations are listed as follows,
\begin{eqnarray}
\left(
\frac {v_k-{i\over 2}}{v_k+{i\over 2}}\right) ^N
&=&\prod _{j=1}^{m^{(1)}}\frac {\mu _j-v_k-{i\over 2}}
{\mu _j-v_k+{i\over 2}}                                
\prod _{l=1,\not= k}^m\frac {v_k-v_l-i}{v_k-v_l+i},
\label{bethe1}
\\
\prod _{j=1}^m\frac {\mu _k-v_j+{i\over 2}}{\mu _k-v_j-{i\over 2}}
&=&\prod _{j=1}^{m^{(1)}}
\frac {\mu _j-\mu _k-i}{\mu _k-\mu _j-i}
\prod _{l=1}^{m^{(2)}}
\frac {\mu _k-\nu _l-{i\over 2}}{\nu _l-\mu _k-{i\over 2}},
\label{bethe2}
\\
\prod _{j=1}^{m^{(1)}}\frac {\nu _k-\mu _j-{i\over 2}}
{\mu _j-\nu _k-{i\over 2}}
&=&
\prod _{j=1,\not= k}^{m^{(2)}}
\frac {\nu _j-\nu _k+i}{\nu _j-\nu _k-i},
\label{bethe3}
\end{eqnarray}
where $k$ take values from 1 to $m,m^{(1)}$ and $m^{(2)}$
in (\ref{bethe1}),(\ref{bethe2}),(\ref{bethe3}) respectively.
Note that we have already redefined the
parameters $v_j,\mu _j,\nu _j$ by shifting ${i\over 2},i$ and
${{3i}\over 2}$ respectively in the above relations.
According to the definition of Hamiltonian in section 3, we can
find the energy of the Hamiltonian $H^0$ as
\begin{eqnarray}
E=N-\sum _{j=1}^m\frac 1{v_j^2+{1\over 4}}.
\end{eqnarray}

We shall use the notations introduced in ref.\cite{EKS1}.
We denote respectively $N_{\uparrow }$ and
$N_{\downarrow }$ as the number of single electrons with spin
up and spin down, and $N_l$, number of local electron pairs.
As we have pointed out, the reference state is chosen to be
$|\uparrow \downarrow>$, and the vacuum state is the full-filled state.
Therefore, an electron means a hole.
The numbers $m,m^{(1)},m^{(2)}$ appeared in above relations
can be written as $m=N_{\uparrow }+N_{\downarrow }+N_l,
~m^{(1)}=N_{\uparrow }+N_{\downarrow },
~m^{(2)}=N_{\downarrow }$. Since the Hamiltonian
commutes with the Hubbard term $U\sum _jX_j$, this term will
not change the eigenvectors and the Bethe ansatz equations.

\section{Analysis of the Bethe Ansatz Equations and Discussions}
Using the standard method \cite{YT}, we next analyze briefly
the Bethe ansatz
equations. By taking the logarithm of the Bethe ansatz equations, we 
obtain the following set of equations,
\begin{eqnarray}
&&N\Phi(v_k,{1\over 2})-\sum _{l=1}^m\Phi(v_k-v_l,1)
-\sum _{j=1}^{m^{(1)}}\Phi(\mu _j-v_k,{1\over 2})=2\pi I_k,
\\
&&\sum _{l=1}^m\Phi(\mu _{\alpha }-v_l,{1\over 2})
-\sum _{\beta =1}^{m^{(1)}}\Phi(\mu _{\alpha }-\mu _{\beta },1)
-\sum _{a=1}^{m^{(2)}}\Phi(\nu _a-\mu _{\alpha },{1\over 2})
=2\pi J_{\alpha },
\\
&&\sum _{\beta =1}^{m^{(1)}}
\Phi(\mu _{\beta }-\nu _b,{1\over 2})
-\sum _{c=1}^{m^{(2)}}\Phi(\nu _c-\nu _b,1)=2\pi K_b,
\end{eqnarray}
where $\Phi(\lambda ,\alpha )=2arctan(\lambda /\alpha)$.
$I_k, J_{\alpha }$ and $K_b$ are integer or half-odd integer depending
on $m, m^{(1)}$ and $m^{(2)}$.  In the thermodynamic limit
$N\rightarrow \infty $, $v_j,\mu _j,\nu _j$ become continuous
varibles, $I_k,J_{\alpha }$ and $K_b$ are connected with the
distribution functions of roots and holes, the Bethe ansatz
equations are rewritten in the following form
\begin{eqnarray}
&&\sigma (v)+\sigma _h(v)
=K_{1/2}(v)-\int _{-B}^Bdv'K_1(v-v')\sigma (v')
+\int _{-B^{(1)}}^{B^{(1)}}d\mu K_{1/2}(v-\mu )\omega (\mu ),
\nonumber \\
&&\omega (\mu )+\omega _h(\mu )
=\int _{-B}^BdvK_{1/2}(\mu -v)\sigma (v)
-\int _{-B^{(1)}}^{B^{(1)}}d\mu 'K_1(\mu -\mu ')
\omega (\mu ')+
\int _{-B^{(2)}}^{B^{(2)}}d\nu K_{1/2}(\mu -\nu )\tau (\nu ),
\nonumber \\
&&\tau (\nu )+\tau _h(\nu )=
\int _{-B^{(1)}}^{B^{(1)}}d\mu K_{1/2}(\nu -\mu )\omega (\mu )
-\int _{-B^{(2)}}^{B^{(2)}}d\nu 'K_1(\nu -\nu ')\tau (\nu '),
\label{ba}
\end{eqnarray}
where $K_{\alpha }(\lambda )=\frac {\alpha }{\pi (\lambda ^2+\alpha ^2)}$.
These results are completely the same as those for SU(4) spin chain,
see ref.\cite{LMSF1}. The energy is written as
\begin{eqnarray}
E=N-2\pi \int _{-B}^BK_{1/2}(v)\sigma (v)dv.
\end{eqnarray}
The ground state energy per site is obtained as
\begin{eqnarray}
E_0=1-({3\over 2}ln2+{\pi \over 4}).
\end{eqnarray}
Because the coupled integral equations (\ref{ba}) are the same as for
the case of normal SU(4) spin chain, the properties
of the low-lying excitations of the system under consideration
remain to be the same as those in
ref.\cite{LMSF1}.

In conclusion, we have presented in this paper the Bethe ansatz
equations for the SU(4) extension of the Hubbard model.
We have proved that the Hamiltonian is equivalent to the
graded SU(4) permutation operator. Using the graded
algebraic Bethe ansatz method, we have found the eigenvalues of
the transfer matrix.

In this paper, we have only dealt with FFBB grading, and there
are only one matrix representations for the
Fermion and Boson generators. We can consider
other representations as presented in ref.\cite{EKS1} and
different gradings. The $SU(4)$
extension of the Hubbard model is different from
the $SU(2|2)$ case, but we use in this paper
the same 16 generators as in ref.\cite{EKS}
to represent the Hamiltonian. The structure of
the Bethe ansatz eigenvectors for $SU(4)$ case seems to be the same
as the $SU(2|2)$ case. We can also give a detailed calculation
of the thermodynamic Bethe ansatz for the $SU(4)$ case as that
of $SU(2|2)$\cite{EKS1}.

\vskip 1truecm
\noindent {\bf Acknowledgements}:One of the authors,H.F. is
supported by JSPS. He thanks M.Batchelor, J.Links, S.Saito, Y.Umeno,
Y.Z.Zhang and
H.Q.Zhou for useful discussions and communications.


\newpage
\begin{thebibliography}{99}
\bibitem{AZR}P.W.Anderson, Science {\bf 235}, 1196(1987);
F.C.Zhang and T.M.Rice, Phys. Rev.{\bf B37}, 3759(1988).
\bibitem{Yang}C.N.Yang, Phys.Rev.Lett.{\bf 63},2144(1989);
C.N.Yang and S.C.Zhang, Mod.Phys.Lett.{\bf B4},759 (1990);
C.N.Yang, Rev.Mod.Phys.{\bf 34},694(1962).
\bibitem{EKS}F.H.L.Essler, V.E.Korepin and K.Schoutens,
Phys.Rev. Lett.{\bf 68},2960(1992); Phys. Rev. Lett. {\bf 70}, 73 (1993).
\bibitem{BKS}J.de Boer, V.E.Korepin, A.Schadschneider,
Phys.Rev.Lett. {\bf 74}, 789 (1995).
\bibitem{FLR}A.Foerster, J.Links and I.Roditi, J.Phys.{\bf A32},
L441 (1999);\\
H.Q.Zhou, X.Y.Ge, J.Links and M.D.Gould, {\it Integrable Kondo
impurities in one-dimensional extended Hubbard models},
cond-mat/9908036;\\
A.A.Aligia and L.Arrachea, {\it Exact solution of a Hubbard chain
with bond-charge interaction}, cond-mat/9406122;\\
A.Schadschneider, {\it Superconductivity in a exactly solvable
Hubbard model with bond-charge interaction}, cond-mat/9411064.
\bibitem{F}H.Fan, J.Phys.{\bf A32}, L509 (1999).
\bibitem{DR}E.Dagotto and T.M.Rice, Science {\bf 271}, 618(1996).
\bibitem{NT}
Yupen Wang,Phys.Rev.{\bf B60},9236(1999); \\
A.A.Nersesyan and A.M.Tsvelik, Phys.Rev.Lett.{\bf 78},3939 (1997);\\
M.T.Batchelor and M.Maslen, J.Phys.{\bf A32},L377(1999);\\
H.Frahm and A.Kundu, {\it Phase diagram of an exactly solvable
t-J ladder model}, cond-mat/9910104.\\
M.T.Batchelor, J.de Gier, J.Links and  M.Maslen,
{\it Exactly solvable quantum spin ladders associated with the
orthogonal and symplectic Lie algebras},
cond-mat/9911043.\\
J.Links and A.Foerster, {\it Solution of a two leg spin ladder
system}, cond-mat/9911096.
\bibitem{LMSZ}Y.Q.Li, M.Ma, D.N.Shi and F.C.Zhang,
Phys.Rev.Lett.{\bf 81}, 3527 (1998);\\
C.Itoi, S.J.Qin and I.Affleck, {\it Phase digram of a 1 dimensional
spin-orbital model}, cond-mat/9910109.
\bibitem{LMSF1}Y.Q.Li, M.Ma, D.N.Shi and F.C.Zhang,
Phys.Rev.{\bf B60},12781 (1999).
\bibitem{EKS1}F.H.L.Essler, V.E.Korepin and K.Schoutens,
Int.J.Mod.Phys.{\bf B8}, 3205, 3243 (1994);
F.H.L.Essler and V.E.Korepin,
Int.J.Mod.Phys.{\bf B8}, 3243 (1994);
\bibitem{YT}C.N.Yang and C.P.Yang, J.Math.Phys.{\bf 10},1115 (1969);\\
M.Takahashi, Prog.Theor.Phys.{\bf 46},1388 (1971).
\end{thebibliography}
\end{document}